\def\x1{\chi_z}
\def\x2{\chi_{zz}}

\def\w{\omega}
\def\w0{\omega_0}
\def\W{\Omega}
\def\W0{\Omega_0}

\def\NV{NV$^-$}

\documentclass[pra,preprint,fleqn]{revtex4}
\usepackage{graphicx}

\begin{document}
\title{Quantum information processing using frequency control of impurity spins in diamond}

\author{A.M.~Zagoskin$^{1,2}$, J.R.~Johansson$^{2}$, S. Ashhab$^{2}$ and Franco Nori$^{2,3}$}

\affiliation{(1) Department of Physics and Astronomy, The University of British Columbia, Vancouver, B.C., V6T 1Z1 Canada}
\affiliation{(2) Digital Materials Laboratory, Frontier Research System, RIKEN, Wako-shi, Saitama 351-0198, Japan}
\affiliation{(3) Center for Theoretical Physics, Physics Department, Center for the Study of Complex Systems, The University of Michigan, Ann Arbor, MI 48109-1040, USA}

\begin{abstract}
Spin degrees of freedom of charged nitrogen-vacancy (NV$^-$)  centers in diamond have large decoherence times even at room temperature, can be initialized and read out using optical fields, and are therefore a promising candidate for  solid state qubits. Recently, quantum manipulations of NV$^-$- centers using RF fields were experimentally realized. In this paper we show; first, that such operations can be controlled by varying the frequency of the signal, instead of its amplitude, and NV$^-$- centers can be selectively addressed even with spacially uniform RF signals;  second,  that when several \NV - centers are placed in an off-resonance optical cavity, a similar application of classical optical fields provides a controlled coupling and enables a universal two-qubit gate (CPHASE). RF and optical control together promise a  scalable quantum computing architecture.
\end{abstract}


\maketitle

\section{Introduction}

Impurity spins in diamond are among the most promising candidates for   solid state quantum hardware. The so called (negatively charged) nitrogen-vacancy  centers (\NV)  have a low-lying spin triplet state $^3\!A$ with a large decoherence time (up to $\sim 350\:\mu$s) at room temperature, which can be initialized and read out using a strong, spin-conserving optical transition to the excited $^3\!E$ state \cite{Oort1990,Jelezko2004,Hanson2006,Gaebel2006,Hanson2006a,Childress2006}. The coherent manipulation of the $^3\!A$ state and its coupling to spins of $^{13}$C \cite{Jelezko2004,Childress2006} and N \cite{Gaebel2006,Hanson2006a} demonstrated the feasibility of \NV-based quantum devices. Though the direct coupling of different \NV- centers, necessary for a scalable architecture, would require placing them too close to each other (within a few nm),  coupling through an optical mode is possible \cite{Lukin2000,Shahriar2002,Greentree2006} using Stark shifts,  in order to tune the coupling on and off.
(Stark shifts in \NV  ~were observed in bulk response \cite{Kaplyans1970,Davies1980,Redman1992} as well as from individual  centers  \cite{Tamarat2006}.)

The use of local time-dependent fields for selective control is a natural approach, but it is not always easily achieved in the case of microscopic qubits. Here we suggest an approach which would allow us to address specific \NV- centers by tuning to their resonant frequency, which can be made position-dependent by the application of a {\em static} nonuniform magnetic field. We will also show that a similar approach using classical optical fields allows controlled coupling and universal two-qubit gates for \NV- centers in an optical cavity.

\section{Model} An \NV- center is a negatively charged complex of a nitrogen impurity and a neighbouring vacancy. It can be formed as a result of nitrogen implantation in the diamond matrix; in experiments so far, the conversion from N to \NV ~was  achieved with a limited efficiency of about 5$\%$ \cite{Gaebel2006}. It is therefore common to find an  \NV- center close to a nitrogen impurity. Unlike an \NV- center, a nitrogen impurity does not have an electric dipolar moment and does not couple to   optical fields.
Consider such a  N-\NV ~complex \cite{Gaebel2006,Hanson2006a}.  Let us choose the [111]-direction  as the $z$-axis. The magnetic moment (spin 1) of the \NV- center and the eigenvalues of its  $z$-component are $\vec{S}$, $M_z = 0,\pm1$; for the N impurity (spin 1/2) they are denoted respectively by $(1/2)\vec{\sigma}$, $m_z = \pm 1/2$;  $\vec{\sigma}$ is the vector of Pauli matrices.    The   $^3\!A$ ground state of the \NV-center is split by the crystal field, while the ($M_z = \pm1$)-states are degenerate, and the ($M_z=0$)-state becoming the true ground state. The ($M_z=0$)-state is also the state leading to enhanced photoluminescence (through the excitation to $^3E$ and subsequent decay through a metastable level $^1\!A$ (see e.g. \cite{Loubser1978}), allowing an optical readout \cite{Jelezko2004,Gaebel2006,Hanson2006,Hanson2006a}. The external magnetic field along the $z$-axis splits the ($M_z = \pm1$)-states as well as the ($m_z = \pm1/2$)-spin states  of the nitrogen impurity. The Hamiltonian of the system is (in the absence of an electric field)
\begin{eqnarray}
H = H_{NV} + H_N + H_{\rm int},
\end{eqnarray}
where
\begin{eqnarray}
H_{NV} = D(S_z)^2 + \kappa \vec{B}\cdot\vec{S},\\
H_{N} = \frac{1}{2}\kappa \vec{B}\cdot\vec{\sigma} + A \vec{\sigma}\cdot\vec{I}.
\end{eqnarray}
Here    $D = 2.88$ GHz \cite{Reddy1987,Oort1988,Redman1991}, $\kappa = 2.8$ MHz/Gs \cite{Gaebel2006,Hanson2006a}, the hyperfine splitting $A=86$ MHz or $114$ MHz depending on the position of the nitrogen in the lattice \cite{Smith1959}, and $\vec{I}$ is its nuclear spin ($I=1$).
The magnetic dipolar interaction
\begin{eqnarray}
H_{\rm int} = \gamma[ \vec{S}\cdot\vec{\sigma} - 3(\vec{S}\cdot\vec{n})(\vec{\sigma}\cdot\vec{n})] \label{eq_int}
\end{eqnarray}
has a scale $\gamma \approx 6.5$ MHz for the distance between \NV ~and N of 2 nm; $\vec{n}$ is the unit vector in the direction connecting N and \NV.

When $B_z=B_{\rm res} = 514$ Gs, the transition  $(M_z=0)\leftrightarrow(M_z=-1)$ in the \NV- center is in resonance with the transition $(m_z=+1/2)\leftrightarrow(m_z=-1/2)$ in N. The term (\ref{eq_int}) then induces coherent transitions in the system, which were experimentally observed in \cite{Gaebel2006,Hanson2006a}. Other resonances,  shifted by $\sim 15$ Gs to either side due to the hyperfine interaction in N, were also observed. (The hyperfine splitting in the \NV- center was too small to be resolved \cite{Hanson2006a}; we neglect it here.)

In order to distinguish different \NV- centers, we now consider the application of a $B_z$-field gradient. For example, if they are placed $10\: \mu$m apart, a field gradient of 1 T/cm will produce a 30 MHz difference in the $(M_z=0) \leftrightarrow (M_z=-1)$  transition frequency between the neighbouring  centers, which is enough for our purposes, as we shall see below.

\section{RF control of single-spin rotations in \NV- centers} If the field $B_z \neq B_{\rm res}$, the transitions $(M_z=0; m_z=1/2) \leftrightarrow (M_z=-1; m_z=-1/2)$ are suppressed.  (Note that these energy conserving transitions do not conserve spin.)
 Single-qubit operation on a  \NV- center can be then performed by applying a {\em spatially uniform} AC field along the $y$-axis, with the resonance frequency $\omega_y = D - \kappa B_z$ (of order of 1.5 GHz) corresponding to the $(M_z=0) \leftrightarrow (M_z=-1)$-transitions:
\begin{equation}
H_{y} = \kappa B_y \cos \omega_y tS_y.
\end{equation}
Due to the $B_z$-gradient,  this frequency is different for different \NV- centers, and  we  have a {\em frequency-based} control.  First, we go to the interaction representation:
\begin{equation}
H_{y} \to U_{NV}H_yU_{NV}^{\dag}, \label{eq_Unitary0}
\end{equation}
where
\begin{eqnarray}\label{eq_UNV}
U_{NV} = \exp[iH_{NV}t] = (1-S_z^2)+(S_z^2\cos \kappa B_zt + i S_z \sin \kappa B_zt)\exp{[iDt]}.
\end{eqnarray}
After dropping the fast rotating terms (rotating wave approximation, RWA), we obtain the effective Hamiltonian
\begin{equation}
 H_{y,{\rm eff}} = \frac{\kappa B_y}{2} \left[ \frac{S_y - [S_z,S_y]_+}{\sqrt{2}}\right] \equiv \frac{\kappa B_y}{2} \sigma^{NV}_y. \label{eq_HeffY}
\end{equation}
The operator in brackets acts as the Pauli matrix $\sigma_y$ on
the subspace $\{M_z=0, M_z=-1\}$. The AC field produces relatively
fast rotations of a chosen \NV- center, in excess of 1 MHz/G.

The use of frequency instead of amplitude RF control is not dictated by the fact that the latter would require a local (within few microns) application of RF fields. The latter is feasible  \cite{Lidar2004} and, on a little larger scale, is being done in experiments with superconducting flux and phase qubits on a regular basis (see e.g. \cite{You2005a}). Nevertheless the frequency control \cite{Liu2006} has certain advantages including, but not limited to, less complex circuitry.

\section{RF control of \NV-N coupling} 
\begin{figure}[t]
\includegraphics[width=8.0cm]{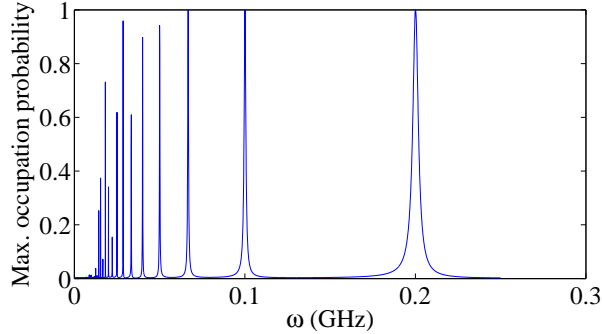}
\caption{\label{fig1a} Maximum occupation probability for the
\NV- center ground state, max $\rho_{00}^{NV}$, when initially in
the excited state, as a function of the AC field frequency,
$\omega$, for the \NV-N detuning $\delta\omega = 0.2$ GHz and the
AC field amplitude $\kappa B_z' = 67$ MHz (corresponding to $\eta
= 0.05$, with static field amplitude $\kappa B_z = 1.34$ GHz);
$\gamma = 6.5$ MHz.  Note additional peaks at integral fractions
of $\delta\omega$, which correspond to the higher-order
(multi-photon) processes, dropped from Eq.~(\ref{eq_Heff2}).}
\end{figure}

\begin{figure}[t]
\includegraphics[width=8.0cm]{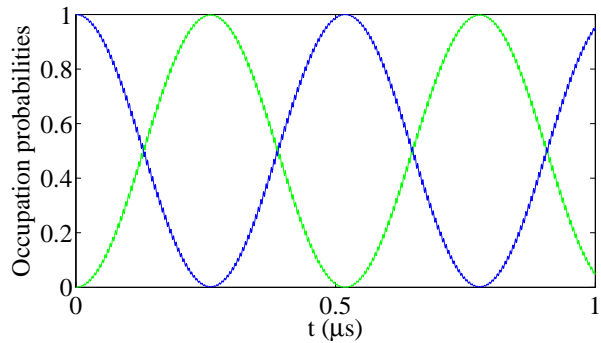}
\caption{\label{fig1b} (Color online) The occupation probabilities of the \NV- center ground state $M_z=0$
(dashed line in blue) and the exited state $M_z=-1$ (in green) as a function
of time, for the same choice of parameters as in Fig.~\ref{fig1a}, and $\omega = 0.2$ GHz. The transition frequency ($\approx 2$ MHz) agrees with the RWA value following from Eq.~(\ref{eq_Heff2}).  }
\end{figure}

A relatively weak AC field with appropriate frequency can turn on the transitions $(M_z=0; m_z=1/2) \leftrightarrow (M_z=-1; m_z=-1/2)$. To see this, we again perform a unitary  transformation of the Hamiltonian to the interaction representation,
\begin{equation}
H \to U_NU_{NV}HU_{NV}^{\dag}U_N^{\dag} -
iU_NU_{NV}\frac{\partial}{\partial
t}\left[U_{NV}^{\dag}U_N^{\dag}\right], \label{eq_Unitary1}
\end{equation}
where $U_{NV}$ was defined in (\ref{eq_UNV}), and
\begin{eqnarray}
U_N = \exp[iH_Nt] = \cos\left[(\kappa B_z + AI_z)t/2\right] + i\sigma_z \sin\left[(\kappa B_z + AI_z)t/2\right].
\end{eqnarray}
The resonance condition for the transition $(M_z=0; m_z=1/2)
\leftrightarrow (M_z=-1; m_z=-1/2)$ is $2\kappa B_z + AI_z - D =
0.$ Assuming a detuning $\delta\omega$ from resonance, i.e.
\begin{equation}
2\kappa B_z + AI_z - D = \delta\omega,
\end{equation}
we find (from Eq. (\ref{eq_Unitary1})) the effective interaction
\begin{equation}
H_{\rm eff} = \gamma(1-3n_z^2)S_z\sigma_z +
\frac{\cos(\delta\omega t)}{2\sqrt{2}} \gamma(2-3n_x^2-3n_y^2)
\left[\sigma^{NV}_x \sigma_x + \sigma^{NV}_y \sigma_y\right] +
h(t)\label{eq_Heff}
\end{equation}
(with the same notation as in Eq.(\ref{eq_HeffY})). Here  $h(t)$ denotes the rest of the terms, which all are fast rotating and should be dropped in the RWA. So is the second term in the r.h.s. of (\ref{eq_Heff}), unless the detuning $\delta\omega$ can be compensated. This is done by an additional field along the $z$-axis,
 $B_z'(t) = \eta \kappa B_z \sin \omega t$. (The corresponding term in the Hamiltonian is not   affected by the transformation (\ref{eq_Unitary1})). After one more unitary transformation,
\begin{equation}
H_{\rm eff} \to U'H_{\rm eff}U'^{\dag} -
iU'\frac{\partial}{\partial t}\left[U'^{\dag}\right],
\label{eq_Unitary2}
\end{equation}
with
\begin{eqnarray}
U = \exp\left[-i \eta\:\kappa\: B_z(S_z+\frac{1}{2}\sigma_z)
\cos(\omega t)/\omega\right],
\end{eqnarray}
 and assuming $\omega = \delta\omega$, we obtain, in the RWA, the following Hamiltonian:
\begin{equation}
\tilde{H}_{\rm eff} = \gamma(1-3n_z^2)S_z\sigma_z +
\frac{J_1(2\eta\kappa B_z/\omega)}{2\sqrt{2}}
\gamma(2-3n_x^2-3n_y^2) \left[\sigma^{NV}_x \sigma_x +
\sigma^{NV}_y \sigma_y\right]. \label{eq_Heff2}
\end{equation}
Otherwise the effective coupling is zero. Here $J_1$ is a Bessel function.  This result, frequency-controlled \NV-N coupling, is confirmed numerically (Figs.~\ref{fig1a},\ref{fig1b}).

The coupling strength in resonance, $\gamma  \approx 6.5$ MHz,
determines how far we should be off resonance in order to achieve
decoupling. This frequency is large enough to justify the
approximations leading to (\ref{eq_Heff2}). The coupling  is
switched on by an AC field $B_z'(t)$. The coupling strength will
be smaller than $\gamma$, but not drastically. Even choosing
$\delta\omega = 0.2$ GHz, $\eta = B_z'/B_{\rm res} = 0.05$, and
remembering that   $\kappa B_{\rm res} \approx 1.5$ GHz, we find
the attenuation of the coupling strength $\approx 0.32$. It will
still produce coherent transitions at $\sim 2$ MHz, which is fast
compared to the decoherence times at room temperature as high as
0.35 ms.

As an aside, in the same way, one can show that when the system is in resonance, an AC field $B_z'(t)$ will suppress the transitions and freeze the spins in \NV ~and N (coherent destrustion of tunneling, see e.g. \cite{Ashhab2007}). This is not useful to the coupling scheme we are describing here.

In order to perform arbitrary one-qubit rotations of nitrogen spins, as well as two-qubit gates between \NV ~and N, we only need to   initialize \NV ~and perform single-qubit rotations on it (see e.g. \cite{Zagoskin2006}, where the phase qubit is  analogous to an \NV- center, and the quantum two-level system to a N impurity spin), which can be done using the RF control (Eq.~(\ref{eq_HeffY})).

\section{ \NV-\NV and indirect N-N couplings}
The scalability of the design requires the coupling between different \NV- centers or \NV-N complexes. For macroscopic qubits this is done through the magnetic flux or charge coupling to the cavity modes \cite{Makhlin2001,Blais2003,Wallraff2004,Zagoskin2006}. In our case, unfortunately, the magnetic coupling is way too weak. Instead, we can use an optical cavity mode and two classical laser fields, along the lines of Refs.~\onlinecite{Imamoglu1999,Shahriar2002}.  This has the disadvantage of involving the $^3\!E$ state, where the decoherence rate is higher. On the other hand, the laser fields are easier to apply locally. By tuning the frequency of the laser field, the interaction strength can be controlled.

Consider two \NV- centers placed in an off-resonance optical cavity. The Hamiltonian of the system is
\begin{eqnarray}
H & =    H_0 + H_{\rm field} + H_{\rm cavity} ,  
\end{eqnarray}
\begin{eqnarray}
H_0  =   \omega_c\: a^{\dag}a &  + \frac{E}{2}(I + \sigma^z_1) + \left(\frac{E+\Omega_1}{2}I +
 \frac{E-\Omega_1}{2}\tau^z_1\right) +  
 \frac{E}{2}(I + \sigma^z_2) + \left(\frac{E+\Omega_2}{2}I + \frac{E-\Omega_2}{2}\tau^z_2\right),
 \end{eqnarray}
 \begin{eqnarray}
 H_{\rm field}  =   H_{\rm field,1} + H_{\rm field,2} = 
   & g_{1x}^{(0)}{\cal E}_1^{(0)} \cos(\omega_1^{(0)}t) \sigma^x_1 + g_{1x}^{(-1)}{\cal E}_1^{(-1)} \cos(\omega_1^{(-1)}t) \tau^x_1 \nonumber \\
  & + 
g_{2x}^{(0)}{\cal E}_2^{(0)} \cos(\omega_2^{(0)}t) \sigma^x_2 + g_{2x}^{(-1)}{\cal E}_2^{(-1)} \cos(\omega_2^{(-1)}t) \tau^x_2,
\end{eqnarray}
\begin{eqnarray}
H_{\rm cavity}    = H_{\rm cavity,1} + H_{\rm cavity,2} = 
    & g_{1z}^{(0)}\epsilon (a^{\dag}+a) \sigma^z_1  + g_{1z}^{(-1)}\epsilon (a^{\dag}+a) \tau^z_1 \nonumber\\ & + 
g_{2z}^{(0)}\epsilon (a^{\dag}+a) \sigma^z_1 + g_{2z}^{(-1)}\epsilon (a^{\dag}+a) \tau^z_2.
\end{eqnarray}
Here the operators
\begin{eqnarray*}
\sigma^z_j = |E_j\rangle\langle E_j| - |(A; M_z=0)_j\rangle\langle (A; M_z=0)0_j|, \\ \tau^z_j = |E_j\rangle\langle E_j| - |(A; M_z=-1)_j\rangle\langle (A; M_z=-1)_j|, \\
\sigma^x_j = |E_j\rangle\langle (A; M_z=0)_j| + |(A; M_z=0)_j\rangle\langle E_j|, \\ 
\tau^z_j = |E_j\rangle\langle (A; M_z=-1)_j| + |(A; M_z=-1)_j\rangle\langle E_j|
\end{eqnarray*}
account for the optical transition $^3\!A \leftrightarrow ^3\!E$ in the $j$th \NV- center.
The term $H_{\rm field}$ describes the interaction of the electric moments of the \NV- centers with the classical laser fields, ${\cal E}^{(0,-1)}_{1,2}$, with frequencies $\omega^{(0,-1)}_{1,2}$, polarized in the $x$-direction; $H_{\rm cavity}$ describes their interaction with the optical cavity mode polarized in the $z$-direction; $\epsilon$ is the ``electric field amplitude for one photon in the cavity".

The idea of the approach remains the same as in the case of  RF control of the \NV-N coupling. For example, in order to induce the transition $(A; M_z=0) \leftrightarrow (E)$ in the \NV- center 1, we switch on the laser field ${\cal E}^{(0)}_{1}$, which is tuned to the frequency $\omega^{(0)}_1 = \omega_c - E$. After performing the unitary transformation with
$ U = U_{\rm cavity} U_0,$ where $U_{\rm cavity} = \exp[i\int^t dt\:\: H_{\rm cavity}]$, and $U_0 = \exp[iH_0t]$, and then a RWA, the resulting term in the Hamiltonian will be,
\begin{equation}
H^{(0)}_{\rm eff,1} = -g^{(0)}_{\rm eff,1} (a^{\dag}\sigma^-_1 + a\sigma^+_1),\:\:
g^{(0)}_{\rm eff,1} = \frac{g_{1z}^{(0)}\epsilon g_{1x}^{(0)}{\cal E}_1^{(0)}}{\omega_c}, \label{eq_cavity}
\end{equation}
and similarly for the rest of the transitions. As in \cite{Imamoglu1999}, the coupling strength is proportional to the classical field amplitude. To target only one \NV- center, we now have two strategies. One is to reproduce our earlier approach and apply a non-uniform electric field. Then, due to the Stark shift, the resonance frequency of a given \NV- center will depend on its location, and the control is realized by applying uniform optical fields at specific frequencies. (Of course, due to the non-uniform static magnetic field applied to the system the ($A\leftrightarrow E$) transition frequencies will already  differ  for different \NV- centers, but the difference is negligible.)

The other strategy is to  apply laser fields locally. Given the transition wavelength 637 nm  this may either put a lower limit on the spacing between \NV- centers, or require e.g. using evanescent modes in waveguides.

The interactions (\ref{eq_cavity}) can produce single-qubit rotations \cite{Imamoglu1999,Shahriar2002}. In our scheme they can be  done more easily with RF pulses. On the other hand, two-qubit gates for different \NV- centers require long-range coupling, which can be achieved through virtual excitations in the cavity. For example, if we only apply the fields ${\cal E}_1^{(0)}$ and ${\cal E}_2^{(0)}$, and eliminate the cavity modes in   linear order by the Schrieffer-Wolff transformation\cite{Schrieffer1966}, we  obtain the effective interaction
\begin{equation}
H^{(00)}_{\rm eff,12} = \frac{2 g^{(0)}_{\rm eff,1}g^{(0)}_{\rm eff,2}}{\omega_c-E} (\sigma^-_1\sigma^+_2 + \sigma^-_2\sigma^+_1). \label{eq_NV-NV}
\end{equation}

We we will also need transitions between the $^3\!A$ levels $M_z=0(-1)$ and the $^3\!E$ state. They can be realized by applying another laser field with  $x$-polarization and at a corresponding resonant frequency $\approx 470$ THz ($\lambda = 637$ nm).   The cavity degrees of freedom or other \NV- centers will not be involved.

Using (\ref{eq_NV-NV}), we can produce, e.g., an entangling transformation
\begin{equation}
(\alpha|M_z=0\rangle_1 + \beta|M_z=-1\rangle_1)\otimes|M_z=0\rangle_2 \to  (\alpha|M_z=0\rangle_1\otimes|M_z=-1\rangle_2 + \beta|M_z=-1\rangle_1\otimes|M_z=0\rangle_2).
\end{equation}
To achieve this, we   perform the following set of operations: \\
1) $\pi$-pulse between $|M_z=0\rangle_1$ and $|E\rangle_1$ ; the result is
$(\alpha|E\rangle_1 + \beta|M_z=-1\rangle_1)\otimes|M_z=0\rangle_2$;\\
2) $\pi$-pulse of the interaction (\ref{eq_NV-NV}), resulting in $(\alpha|M_z=0\rangle_1\otimes|E\rangle_2 + \beta|M_z=-1\rangle_1\otimes|M_z=0\rangle_2)$; \\
3) $\pi$-pulse between $|E\rangle_2$ and $|M_z=-1\rangle_2$; the outcome is  $(\alpha|M_z=0\rangle_1\otimes|M_z=-1\rangle_2 + \beta|M_z=-1\rangle_1\otimes|M_z=0\rangle_2)$, as required.
After enabling the effective \NV-\NV coupling through the cavity, the operations on N impurities coupled to different \NV- centers can be realized in the same way as in Ref.~\cite{Zagoskin2006}.

Although the above procedure nicely demonstrates the possibility
of using the cavity to perform two-qubit gates, it takes any state
of the form $(\alpha|M_z=0\rangle_1 +
\beta|M_z=-1\rangle_1)\otimes|M_z=-1\rangle_2$ outside the
computational basis. Instead, the C-phase gate could be implemented as
follows: \\
1) $\pi$-pulse between $|M_z=0\rangle_1$ and $|E\rangle_1$;\\
2) $2\pi$-pulse of the interaction (\ref{eq_NV-NV});\\
3) $\pi$-pulse between $|M_z=0\rangle_1$ and $|E\rangle_1$.\\
An inspection of the above procedure shows that the three states
$|M_z=0\rangle_1 |M_z=0\rangle_2$, $|M_z=-1\rangle_1 |M_z=0\rangle_2$ and
$|M_z=-1\rangle_1 |M_z=-1\rangle_2$ are left unchanged at the end of
the procedure, whereas the state $|M_z=0\rangle_1 |M_z=-1\rangle_2$
acquires a minus sign (note that we are not including here the
phases accumulated as a result of single-qubit Larmor precession).
This two-qubit gate, along with single-qubit rotations, form a
universal set of gates for quantum computing.

The requirements to the optical cavity are high, but not impossible. In order to resolve $\sim 1.5$ GHz against the $\sim 470$ THz resonant frequency,  the quality factor of the cavity should be at least of order $3\times10^5$. We could somewhat improve the situation by adding one more step and swapping the states $M_z=0$ and $M_z=1$. This can be done again using the RF field $B_y(t)$ (Eq.~(\ref{eq_HeffY})), this time with the frequency $\omega'_y = D + \kappa B_z$ (of order of 4.5 GHz). This increases the difference of the states involved in the optical coupling from $\sim 1.5$ GHz to $\sim 3$ GHz.

\section{Conclusions} We propose a frequency-controlled approach to coherent manipulation of spin states of \NV- centers and \NV-N complexes in diamond. It allows to address different spins through the difference in their resonance frequencies, induced by a static non-uniform magnetic field.  The time-domain manipulations are performed using uniform RF fields. Different \NV- centers and \NV-N complexes can be coupled optically through the virtual excitations in an optical cavity. Here both frequency control with spatially uniform AC fields  and with local AC fields are possible. The required cavity quality factor is high, but achievable. Our results show that small-scale quantum information processing devices based on impurity spins in diamond  may be feasible in the near future.

{\em Acknowledgements.--} We are greatly indebted to A. Greentree and R. Hanson for very helpful remarks and suggestions, and a thorough reading of the manuscript. This work was supported in part by the ARO, LPS, NSA, and by the 
NSF Grant No. EIA-0130383. A.Z. acknowledges partial support
by the NSERC Discovery Grants Program. S.A. was supported by 
the JSPS.

\bibliography{DIAMOND_January2007}
\bibliographystyle{apsrev}

\end{document}